\documentclass[pre,twocolumn,preprintnumbers,amsmath,amssymb,superscriptaddress]{revtex4-2}

\usepackage{color}
\usepackage[pdftex]{graphicx}
\usepackage{verbatim}
\usepackage{amssymb}
\usepackage{amsmath}
\usepackage{color}
\usepackage[export]{adjustbox}
\renewcommand{\figurename}{Fig.}

\newcommand{\blue}[1]
{\textcolor{black}{ #1}}
\usepackage{mathtools}

\usepackage{makecell}
\usepackage{enumerate}
\usepackage{mathtools}
\usepackage{stmaryrd}
\usepackage[toc,page]{appendix}
\usepackage{textcomp}
\usepackage{gensymb}
\usepackage[shortlabels]{enumitem}

\begin{document}
	
	\title{Full Record Statistics of 1d Random Walks}
	\author{L\'eo R\'egnier}
	\address{Laboratoire de Physique Th\'eorique de la Mati\`ere Condens\'ee,
		CNRS/Sorbonne University, 4 Place Jussieu, 75005 Paris, France}
	\author{Maxim Dolgushev}
	\address{Laboratoire de Physique Th\'eorique de la Mati\`ere Condens\'ee,
		CNRS/Sorbonne University, 4 Place Jussieu, 75005 Paris, France}
	\author{Olivier B\'enichou}
	\address{Laboratoire de Physique Th\'eorique de la Mati\`ere Condens\'ee,
		CNRS/Sorbonne University, 4 Place Jussieu, 75005 Paris, France}
	\begin{abstract}
		We develop a comprehensive framework for analyzing full record statistics, covering record counts $M(t_1), M(t_2), \ldots$, and their corresponding attainment times $T_{M(t_1)}, T_{M(t_2)}, \ldots$, as well as the intervals until the next record. From this multiple-time distribution, we derive general expressions for various observables related to record dynamics, including the conditional number of records given the number observed at a previous time and the conditional time required to reach the current record, given the occurrence time of the previous one. Our formalism is exemplified by a variety of stochastic processes, including biased nearest-neighbor random walks, asymmetric run-and-tumble dynamics, and random walks with stochastic resetting.
	\end{abstract}
	\maketitle

	\section{Introduction}
	The statistics of records in the time series $\left( X_t\right)_{t=0,1,\ldots}$ is a longstanding topic in probability theory \cite{Levy:1940,Chandler:1952,Nevzorov:1988,Majumdar:2010b}, finding applications in various fields such as hydrology \cite{Vogel:2001}, finance \cite{Wergen:2011}, climatology \cite{Morit:2022}, and sports \cite{Gembris:2002}. In a time series $(X_t)$, we have a record at time $t$ if $X_t$ exceeds all preceding values $X_{t'}$ with $t'=0,1,\ldots, t-1$. While the topic was first introduced in the context of independent identically distributed random variables $X_t$ 
	\cite{Chandler:1952}, recent works have considered records when $\left( X_t\right)$ represents the successive positions of a random walk (RW), i.e., $X_{t+1}-X_t=\eta_t$, where the $\eta_t$ denote the steps  of the RW.
	Records of RWs have been studied when the steps $\eta_t$ are independent \cite{Majumdar:2008b,Godreche:2014,Majumdar:2010b,Godreche:2017,Benichou:2016a} or correlated \cite{Godreche:2022,Regnier:2023,Godreche:2015}, including cases with resetting of the position $X_t$ \cite{Majumdar:2022,DeBruyne:2021,Evans:2011}.
	The existing results on  record  dynamics are however essentially restricted to the {\it single}-time distribution of the  number $M(t)$ of records reached at time $t$, as well as the joint distribution of both the number of records and the time at which the current running record was achieved. These results have been obtained for Brownian motion, with and without bias \cite{Levy:1940,Majumdar:2010b,Schehr:2010,Buffet:2003}, run-and-tumble particles \cite{Singh:2019,Cinque:2021,DeBruyne:2021},  random acceleration process \cite{Godreche:2022}, fractional Brownian motion \cite{Wiese:2006}, resetting Brownian motion \cite{Evans:2011,Majumdar:2022,Singh:2021} and resetting run-and-tumble particles \cite{Evans:2018,Tucci:2022,Mori:2020,Mori:2022}. For most studies on records, the large time properties are described using the continuous  in time asymptotic limit process ${\rm d}X_t/{\rm d}t=\eta_t$ \cite{Godreche:2022}. The continuum counterpart of the number of records  is then the maximum of the walk, which coincides  with the number of records \blue{as well as with the value (or position) of the record} in the case of nearest neighbor random walks (every time a new record is reached, the maximum position is incremented by one). Characterizing the record dynamics at long times is then equivalent to characterizing the maximum dynamics. 
	
	Since  record dynamics are non-Gaussian (see the single-time distribution \cite{Redner:2001}) and non-Markovian (knowledge of $M(t')$ at time $t'$ is insufficient to determine the properties of $M(t)$ for time $t>t'$, because the position of the random walk at time $t'$ is not known), 
	determining {\it multiple}-time quantities is  essential to fully characterize the process. A first step in this direction has been done in Refs.~\cite{Benichou:2016a,Benichou:2016b}, where  two-time distributions have been determined. However, these works concerned only two-time quantities and were limited to Brownian motion and Brownian bridges. \\
	
	\begin{figure}[t!]
		\centering
		\includegraphics[width=\columnwidth]{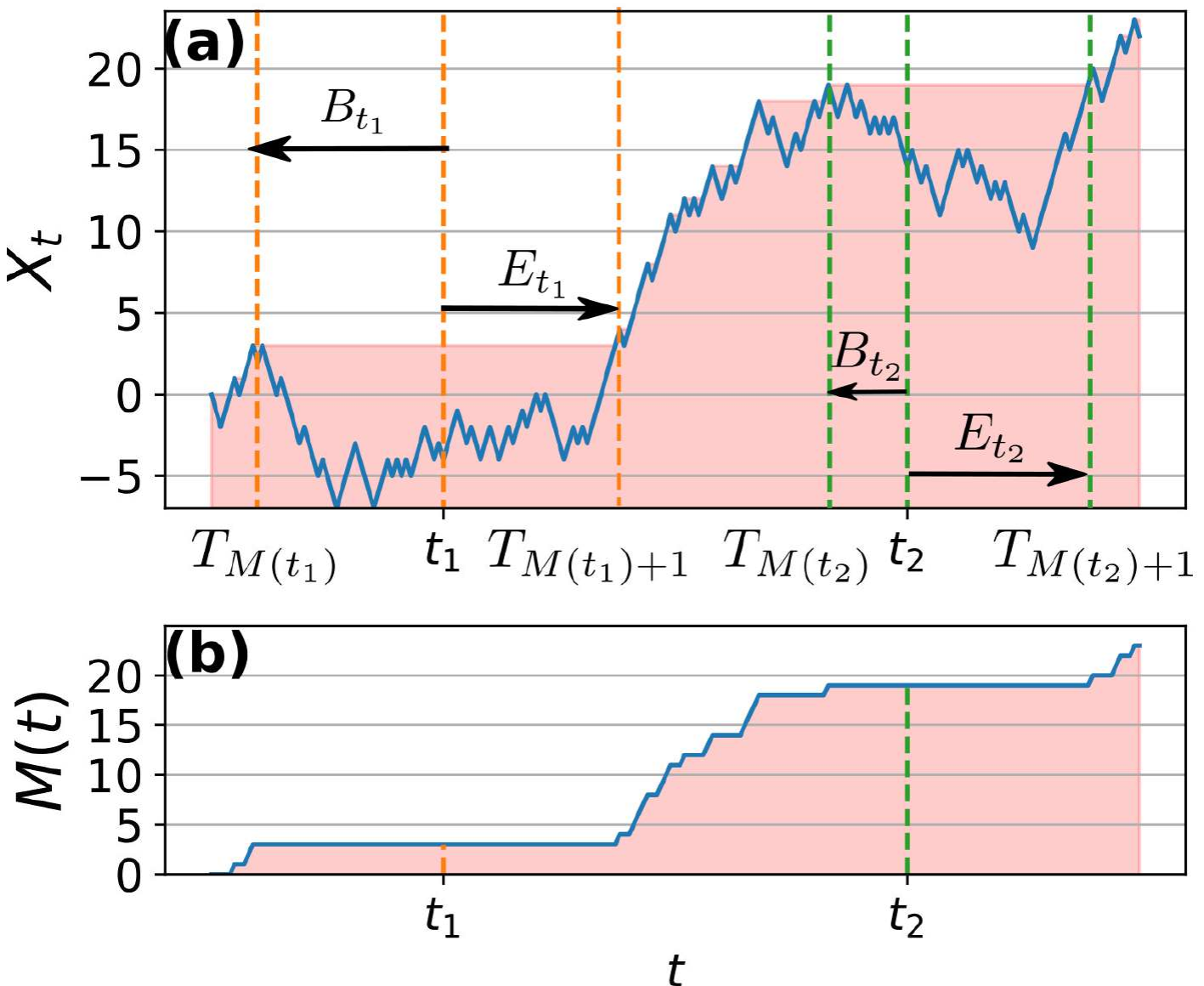}
		\caption{{\bf Representation of the record process.}  (a) Space and time trajectory of a $1d$ symmetric nearest neighbour random walk (RW) and (b) the dynamics of its number of records, which coincides with its maximum. We are interested in the joint statistics of the number of records at different times $M(t_1)$ and $M(t_2)$ as well as the time of occurrence of the last (resp. next) record event, $T_{M(t_1)}=t_1-B_{t_1}$ (resp. $T_{M(t_1)+1}=t_1+E_{t_1}$). }
		\label{fig:Illustration}
	\end{figure}

	\blue{The pathway to the generalization of these results lies in the fact that the times when new maxima are achieved form a Markov process \cite{Morters:2010}: Indeed, at the moment when the maxima $m$ is reached, the position of the RW is known and given by $m$.} 
	%\blue{A key remark to generalize these results is that the times when new maxima are achieved form a Markov process \cite{Morters:2010}: indeed, when the maxima $m$ is reached for the first time, the position of the RW also is known and given by $m$.} 
	Here we develop a general formalism based on renewal theory \cite{Lamperti:1958,Godreche:2001,Godreche:2022}, utilizing the record ages $\tau_k$, defined as the time elapsed between the $k^\text{th}$ and \blue{$(k+1)^\text{th}$} records, as renewal times. Indeed, we rely on the simple but important remark that the number of records  can  be seen as the number of renewal events of length given by  record ages. \blue{Note that our approach connects the number of records of a discrete RW and the maximum of its continuous counterpart. In the continuous case, the number of records is not clearly defined, as it tends to infinity at every time scale \cite{Morters:2010,Benigni:2018}. To establish this connection, we employ a discretization step that determines the additional length the RW must traverse to be considered to have surpassed its previous record. By setting this length scale to unity, the expressions derived for the maximum and number of records align \cite{Godreche:2022} in the asymptotic limit.}
	
	The merits of our approach are the following. (i) It allows us to (re)obtain  very naturally the distributions   of  standard observables such as the number of records at time $t$ or  the time  at which this record was reached (i.e. the time of last record), but also that of the time of next record, which has not been considered yet; (ii) It applies  both to discrete (records) and continuous (maxima) processes; (iii) Importantly, it gives access to {\it all} $n$ time-distributions, \blue{$n$ arbitrary large,} associated to these observables;  (iv) It provides results not limited to Brownian Motion, but which hold for any nearest-neighbor Markovian RW. Explicit expressions for representative examples such that the asymmetric Brownian Motion (aBM) and the asymmetric Run-and-Tumble particle (aRTP) are obtained. (v) It can be extended  to cover the case of non nearest-neighbor RWs, as illustrated by the example of resetting RWs. \\
	\blue{In terms of limitations and potential future directions, it's important to note that our study does not encompass processes characterized by long jumps, such as L\'evy flights \cite{Godreche:2017,Bouchaud:1990}, as the number of records and the maximum position exhibit fundamental differences. Similarly, our analysis does not extend to processes with strong non-Markovian properties, such as fractional Brownian motion \cite{Mandelbrot:1968} or self-repulsive RWs \cite{Amit:1983,Pietronero:1983,Obukhov:1983}, where each step is influenced by the entire past history of the trajectory. }

	The article is structured as follows:
	We start by deriving the single time distribution of the number of records, time of occurrence of the last record, backward record time and forward record time as an illustration of the method. Then, we derive the multiple-time  distributions of the same observables, and provide explicit expressions on the example of the distribution of the times at which the last record events occurred at two different times. Finally, we generalize our approach to aging record times by considering the example of resetting RWs.

	\section{Single-time distributions.}
	We consider a nearest-neighbor Markovian $1d$ lattice RW \cite{Klinger:2021}.  In this case, the running  maximum of the RW and the number of records coincide \cite{Godreche:2022b}. Successive positions are denoted as $\left( X_{t} \right)_{t=0,1,\ldots}$. 
	To introduce our formalism, we first show how to recover the classic asymptotic distribution of the number of records at time $t$, $M(t)$. Our approach relies on the random variable $\tau_k$, defined as the time elapsed between the breaking of record $k$ and $(k + 1)$. The RW being Markovian and nearest-neighbor, the $\tau_k$ are independent identically distributed random variables with probability distribution $F(t)=\mathbb{P}\left(\tau_k=t \right)$, corresponding to the renewal times \cite{Lamperti:1958,Godreche:2001}, as they represent the times between two record events. This simple remark makes $M(t)$  a renewal process. In turn, as we show below, it allows us to use  the powerful methodology of renewal processes to obtain easily the statistical properties of $M(t)$ (including $n$ time quantities).
	\subsection{Number of records}
	We first relate the statistics of $M(t)$ to that of the times $\tau_k$,
	\begin{align}
		\mathbb{P}\left(M(t)\geq m \right)&=\mathbb{P}\left(\tau_0+\ldots+\tau_{m-1}\leq t \right) \nonumber \\
		&=\mathbb{E}\left(H(t-\tau_0-\ldots-\tau_{m-1}) \right) \; ,  \label{eq:Mt_av}
	\end{align}
	where $H$ is the Heaviside function \blue{and $\mathbb{E}(\ldots)$ the average operator}. This equality states that at least $m$ records have been reached by time $t$ if and only if there were $m$ record events before time $t$. To compute the distribution of $M(t)$, we use the discrete Laplace transform of a 
	\blue{of a probability distribution function $f(\tau)$ of the random variable $\tau$ (having realizations $t=0,1,2,\dots$):}
	%function $f(t)$, \blue{ eventually the probability distribution of a positive random variable $\tau$}, 
	\begin{align}
		\label{eq:Lap}
		\mathcal{L}\lbrace f(t) \rbrace=\hat{f}(s)\equiv \sum_{t=0}^\infty  f(t)e^{-st}=\mathbb{E}\left(e^{-s\tau} \right) \; .
	\end{align}
	
	Starting from the distribution $F(t)$ of record ages $\tau_k$, we derive the Laplace transform expression for the distribution of the number $M(t)$ of records at time $t$ by Laplace transforming Eq.~\eqref{eq:Mt_av},
	\begin{align}
		&\mathcal{L}\lbrace\mathbb{P}\left(M(t)\geq m \right)\rbrace \nonumber \\
		&=\mathbb{E}\left( \sum_{t=0}^\infty e^{-st} H(t-\tau_0-\ldots-\tau_{m-1}) \right) \nonumber \\
		&=\mathbb{E}\left( \sum_{t=\tau_0+\ldots+\tau_{m-1}}^\infty e^{-st} \right) \nonumber \\
		&=\frac{1}{1-e^{-s}}\mathbb{E}\left(e^{-s(\tau_0+\ldots+\tau_{m-1})} \right) \nonumber \\
		&=\frac{1}{1-e^{-s}}\hat{F}(s)^m \; . \label{eq:Max_t}
	\end{align}
	In particular, Eq.~\eqref{eq:Max_t} provides a simple expression for the moments of the number of records\blue{, given by:
		\begin{align}
			\mathcal{L}\lbrace \mathbb{E}\left( M^k(t) \right) \rbrace &= \sum_{m=0}^\infty m^k \mathcal{L}\lbrace\mathbb{P}\left(M(t)= m \right)\rbrace  \; .
		\end{align}
		Then, considering the large time, small $s$, limit for which $\hat{F}(s)$ approaches one (see Eq.~\eqref{eq:Lap} with $s=0$), one can replace the sum by an integral to obtain the following simple expression:
		\begin{align}
			\mathcal{L}\lbrace \mathbb{E}\left( M^k(t) \right) \rbrace &\sim \int_0^\infty {\rm d}m m^k (-\partial_m)\mathcal{L}\lbrace\mathbb{P}\left(M(t)\geq m \right) \rbrace \nonumber \\
			&\sim \frac{\Gamma(k+1)}{s(-\ln \hat{F}(s))^k} \; .
	\end{align} }
	Finally, obtaining records statistics is reduced to the determination of the statistics of the $\tau_k$, or in other words, the solving of a first-passage time (FPT) problem from the origin to a point at distance one.
	In Appendix \ref{app:FPT}, we remind for self-consistency  how to obtain these FPT distributions for various models, such as the aBM as well as the aRTP in both discrete and continuous time (see also \cite{Klinger:2022}). It results in the following expressions in the large time, small $s$, limit \cite{Redner:2001,Hughes:1995,DeBruyne:2021,Tucci:2022,Lopez:2014}:
	\begin{align}
		-\ln \hat{F}(s)\sim w(s) \equiv \begin{cases}
			\sqrt{s/D} \; \text{(BM)} \\
			-\frac{\mu}{2D}+\sqrt{\frac{\mu^2}{4D^2}+\frac{s}{D}}  \; \text{(aBM)} \\
			\sqrt{\frac{s^2}{v^2}+\frac{s}{v^2T}} \; \text{(RTP)} \\
			\frac{\sqrt{4sTv^2(sT+1)+\mu^2}-\mu(2sT+1)}{2T(v^2-\mu^2)} \; \text{(aRTP)}
		\end{cases}\label{eq:ws}
	\end{align}
	where $D$ is the diffusive constant, $\mu$ is the bias, $T$ is the tumbling rate, and $v$ is the average absolute speed during a run of the RTP (see Appendix \ref{app:FPT} for precise definition of these models and extension of the aRTP with state-dependent tumbling rate). As we proceed to show, numerous record-type observables have generic and simple expressions as functions of $w(s)$. 
	
	\subsection{Time of last record}
	Using renewal theory, one can easily obtain the joint distribution of the number $m$ of records at time $t$ and the time $T_m=\tau_0+\ldots+\tau_{m-1}$ at which this last record was reached, as well as the backward (forward) record times $B_t\equiv t-T_{M(t)}$ (resp. $E_t\equiv T_{M(t)+1}-t$) corresponding to the duration of the current record (resp. time left until observation of a new record), represented in Fig.~\ref{fig:Illustration}. We note that, while the time at which the last record is reached is a standard observable (see for instance \cite{Singh:2019,Singh:2021,Majumdar:2010,Majumdar:2010b}), the forward record time has not been studied  in the context of records.
	
	To obtain the joint distribution of $(M(t),T_{M(t)})$, we use that having observed $M(t)=m$ records at time $t$ with the last record occurring at time $T_{M(t)}=T_m=y$ is exactly the event of observing $m$ records at time $y$ smaller than $t$ and observing the \blue{$(m+1)^\text{th}$} record after $t$ (in other words, $T_{m+1}>t$). By taking the Laplace transform with respect to both $t$ 
	and $y$ of this event, we obtain that:
	\begin{align}
		&\mathcal{L}\lbrace \mathbb{P}\left( T_m=y,M(t)=m \right) \rbrace \nonumber \\
		&=\sum_{t,y=0}^\infty e^{-uy-st}\mathbb{E}\left( H\left(T_{m+1}-t \right)H(t-y)\delta\left(y-T_m \right)\right) \nonumber \\
		&=\frac{1}{1-e^{-s}}\mathbb{E}\left( e^{-(s+u)(\tau_0+\ldots+\tau_{m-1})}(1-e^{-s\tau_{m}})\right) \nonumber \\
		&=\hat{F}(s+u)^m\frac{1-\hat{F}(s)}{1-e^{-s}} \label{eq:1_time_record_discrete}  \\
		& \sim e^{-w(s+u)m}\frac{w(s)}{s} \label{eq:1_time_record}
	\end{align}
	While Eq.~\eqref{eq:1_time_record_discrete} corresponds to the exact expression of the number of records at any time for the discrete process, Eq.~\eqref{eq:1_time_record} corresponds to the large time asymptotic limit which also describes exactly the maximum of the continuous asymptotic process (see Appendix \ref{App:discr_cont} for details). Eq.~\eqref{eq:1_time_record} provides the Laplace transform of the joint distribution of the maximum and time to reach of the maximum which has been studied for a variety of continuous stochastic processes \cite{Majumdar:2008c,Mori:2022,Buffet:2003,Singh:2019,Wiese:2006}. 
	
	\subsection{Backward record time}
	Then we compute the joint distribution of the backward record time and number of records, $(B_t,M(t))$. To do so, we consider the event of having reached $M(t)=m$ records with the time elapsed since the last record $B_t=t-T_m=y$. This event corresponds to having reached $m$ records at time $t-y$ and not having reached any new record in the time $y$ left (with $y\leq t$). In other words, 
	\begin{align}
		&\mathcal{L} \lbrace \mathbb{P}\left( B_t=y,M(t)=m \right)\rbrace \nonumber  \\
		&=\sum_{t,y=0}^\infty e^{-uy-st}\mathbb{E}\left( H\left(T_{m+1}-t \right)H(t-y)\delta\left(t-y-T_m \right)\right) \nonumber \\
		&=\frac{1}{1-e^{-s-u}}\mathbb{E}\left( e^{-s(\tau_0+\ldots+\tau_{m-1})}(1-e^{-(s+u)\tau_{m}})\right) \nonumber
	\end{align}
	\begin{align}
		\mathcal{L} \lbrace \mathbb{P}\left( B_t=y,M(t)=m \right)\rbrace &=\hat{F}(s)^m\frac{1-\hat{F}(s+u)}{1-e^{-s-u}} \label{eq:1_Backward_record_discrete}  \\
		&\sim e^{-w(s)m}\frac{w(s+u)}{s+u} \label{eq:1_Backward_record}
	\end{align}
	Once again, Eq.~\eqref{eq:1_Backward_record_discrete} is the exact discrete expression and Eq.~\eqref{eq:1_Backward_record} is the asymptotic continuum limit  (see Appendix \ref{App:discr_cont} for details). It is noteworthy that the random variables $T_{M(t)}$ and $B_t$ essentially represent the same observables; however, the former is commonly utilized in extreme value statistics \cite{Singh:2021,Singh:2022,Majumdar:2010,Majumdar:2010b}, while the latter is more prevalent in renewal theory \cite{Godreche:2001,Losidis:2023}. 
	
	One can perform explicit inverse Laplace transforms of Eqs. \eqref{eq:1_time_record} and \eqref{eq:1_Backward_record} (see Appendix \ref{app:C_expressions}) to obtain the expressions derived in \cite{Singh:2019,Majumdar:2010b,Buffet:2003,Majumdar:2008c,Majumdar:2004} for the specific cases of  aBM and RTP. We stress that these results were obtained by a different method, relying on path decomposition technique. 
	
	\subsection{Forward record time}
\begin{figure}[t!]
		\includegraphics[width=\columnwidth]{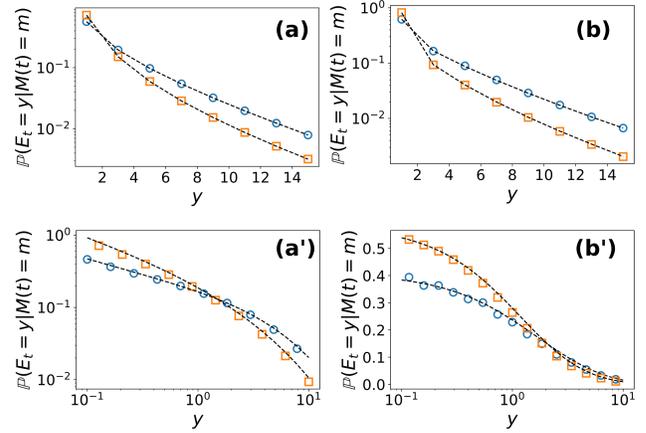}
		\caption{{\bf Conditional forward record time.} Distribution of $E_t$  conditioned on the number of records $M(t)=2$ (blue circles) or $8$ (orange squares), for the asymmetric walks at time $t=10$ for \textbf{(a)} the discrete aBM (probability to go to the right $3/4$) \textbf{(b)} the discrete aRTP (probability to go to the right coming from the right $7/8$, and to the left coming from the left $3/8$) \textbf{(a')} aBM (diffusion constant $D=1$ and bias $\mu=0.5$) \textbf{(b')} aRTP (tumbling rate $T=1/2$, average absolute speed $v=1$ and bias $\mu=0.5$), and $t=10$. The black dashed lines in \textbf{(a)} and \textbf{(b)} stand for the series expansion of the discrete expression in Eq.~\eqref{eq:forward_record_Laplace_discrete}. The black dashed lines in \textbf{(a')} and \textbf{(b')} stand for the expressions of Eqs. \eqref{eq:Forward_aBM_t} and \eqref{eq:Forward_aRTP_t}.}
		\label{fig:Et}
	\end{figure}

	Finally, we compute the joint distribution of the forward record time and current number of records, $(E_t,M(t))$. To do so, we consider the event of reaching $M(t)=m$ records with the time until the next one $E_t=T_{m+1}-t=T_m+\tau_m-t=y$. This event corresponds to achieving $m$ records at a time less than $t$ and $m+1$ records at exactly $t+y$. Thus, 
	\begin{align}
		&\mathcal{L} \lbrace \mathbb{P}\left( E_t=y,M(t)=m \right) \rbrace \nonumber  \\
		&=\sum_{t,y=0}^\infty e^{-uy-st}\mathbb{E}\left( \delta(y+t-T_{m}-\tau_m)H(t-T_m) \right) \nonumber \\
		&=\mathbb{E}\left( e^{-s(\tau_0+\ldots+\tau_{m-1})} \frac{e^{-u\tau_m}- e^{-s\tau_m}}{1-e^{-s+u}} \right) \nonumber \\
		&=\hat{F}(s)^m \frac{\hat{F}(u)-\hat{F}(s)}{1-e^{-s+u}} \label{eq:forward_record_Laplace_discrete} \\
		&\sim e^{-w(s)m}\frac{w(s)-w(u)}{s-u} \; , \label{eq:forward_record_Laplace} 
	\end{align}
	where Eq.~\eqref{eq:forward_record_Laplace_discrete} is the exact discrete expression and Eq.~\eqref{eq:forward_record_Laplace} its asymptotic continuum limit. This general result can be illustrated by the explicit examples of the aBM and the aRTP (see Fig.~\ref{fig:Et}). In the continuum setting, one can Laplace inverse Eq.~\eqref{eq:forward_record_Laplace}, and get for the aBM the following joint distribution:
	\begin{align}
		&\mathbb{P}\left(M(t)=m , E_t=y \right)=\int_0^t {\rm d}t' \frac{m e^{-\frac{\mu^2(y+t')}{4D}-\frac{(m-\mu(t-t'))^2}{4D(t-t')}}}{4D\pi [(y+t')(t-t')]^{3/2}} \; ,
		\label{eq:Forward_aBM_t}
	\end{align}
	and for the aRTP:
	\begin{align}
		&\mathbb{P}\left(M(t)=m , E_t=y \right) \nonumber \\
		&=\int_0^t  \frac{(2T)^{-1}{\rm d}t'}{t-t'+y}\frac{e^{-(t+y)/2T}}{\sqrt{v^2-\mu^2}}I_1\left((t-t'+y)\frac{\sqrt{v^2-\mu^2}}{2vT} \right) \nonumber \\
		&\times \Bigg[\delta\left(t'-\frac{m}{v+\mu} \right)+\frac{m(2T)^{-1}H(t'-m/(v+\mu))}{\sqrt{(t'(v-\mu)+m)(t'(v+\mu)-m)}} \nonumber \\
		&\times I_1\left(\sqrt{(t'(v-\mu)+m)(t'(v+\mu)-m)}/(2vT)\right) \Bigg].
		\label{eq:Forward_aRTP_t}
	\end{align}
	Here $T$, $\mu$ and $v$ are as defined in Eq.~\eqref{eq:ws}.

	We represent these two analytical expressions in Fig.~\ref{fig:Et} as well as their discrete counterpart. We compare the distributions conditioned on the number $M(t)=m$ of records at time $t$ for different values of $m$. In particular, we observe that having a high number of records increases the probability of observing the next record in a short amount of time.  This highlights the importance of correlations in  records processes, as well as the necessity to quantify   multiple-time distributions to fully understand records dynamics. This is the object of the next section.

	\section{Multiple-time point distributions}
	In this section, we show that  the methodology introduced above allows us  to derive straightforwardly multiple-time distributions. For simplicity, we focus on the large time and large number of records asymptotics, even though our method gives the exact results at all times \blue{for the description of the maximum of the continuous process}, as shown in the previous section. In particular, we use continuous Laplace transforms and the record ages of the corresponding continuous processes. 
	
	\subsection{Two-time distribution}
	We start by the two-time distribution of the number of records, before generalizing it to an arbitrary number of time points and considering other record-type observables.
	
	As for the single-time case, we make use of the record ages $\tau_k$, as 
	\begin{align}
		&\mathbb{P}\left( M(t_1) \geq m_1, M(t_2)\geq m_2 \right) \nonumber \\
		&=\mathbb{P}\left( \tau_0+\ldots+\tau_{m_1-1}\leq t_1, \tau_{0}+\ldots +\tau_{m_2-1} \leq t_2 \right) \nonumber \\
		&=\mathbb{E}\left(H(t_1-\tau_0-\ldots-\tau_{m_1-1})H(t_2-\tau_0-\ldots-\tau_{m_2-1}) \right) \; .
	\end{align}
	By Laplace transforming with respect to both time variables $t_1$ and $t_2$, we obtain ($m_1\leq m_2$)
	\begin{align}
		&\mathcal{L}\lbrace \mathbb{P}\left( M(t_1) \geq m_1, M(t_2)\geq m_2 \right) \rbrace  \nonumber \\
		&=\iint_{0}^\infty {\rm d}t_1 {\rm d}t_2 e^{-s_1t_1-s_2t_2}\mathbb{P}\left( M(t_1) \geq m_1, M(t_2)\geq m_2 \right) \nonumber\\
		&=\frac{\mathbb{E}\left(e^{-(s_1+s_2)\left(\tau_0+\ldots+\tau_{m_1-1}\right)-s_2\left(\tau_{m_1}+\ldots+\tau_{m_2-1}\right)}  \right) }{s_1s_2}\nonumber \\
		& = \frac{e^{-w(s_2+s_1)m_1-w(s_2)(m_2-m_1)}}{s_1s_2}  \; . \label{eq:M1_M2_cum}
	\end{align}
	Thus, by deriving Eq.~\eqref{eq:M1_M2_cum} for $m_1<m_2$, we get the density 
	\begin{align}
		&\mathcal{L}\lbrace \mathbb{P}\left( M(t_1) = m_1, M(t_2) = m_2 \right) \rbrace  \nonumber \\
		&=\frac{e^{-w(s_2+s_1)m_1-w(s_2)(m_2-m_1)}}{s_1s_2}w(s_2)(w(s_1+s_2)-w(s_2)) \; .
	\end{align}
	Additionally, one gets the Laplace transform of the probability of having a maximum $M(t_1)$ of value $m$ at time $t_1$ and $M(t_2)$ of same value $m$ at time $t_2$,
	\blue{
		\begin{align}
			&\mathcal{L}\lbrace \mathbb{P}\left( M(t_1)=M(t_2)=m \right) \rbrace \nonumber  \\
			&=\mathcal{L}\lbrace \mathbb{P}\left( M(t_1)=m \right) \rbrace \nonumber \\
			& \; \;  -\int_{m_2 > m}  {\rm d}m_2 \mathcal{L}\lbrace \mathbb{P}\left( M(t_1)=m,M(t_2)=m_2 \right) \rbrace \nonumber \\
			& \; \;  -\int_{m>m_2}  {\rm d}m_2 \mathcal{L}\lbrace \mathbb{P}\left( M(t_2)=m_2,M(t_1)=m \right) \rbrace \nonumber \\
			&=\frac{e^{-w(s_2+s_1)m}}{s_1s_2}\left(w(s_1)+w(s_2)-w(s_1+s_2) \right) \; . \label{eq:Equality_mk}
		\end{align}
	}
	\blue{Of note, here the Laplace transform involves both $t_1$ and $t_2$, thus the Laplace transform of $\mathbb{P}\left( M(t_1)=m \right)$ includes an additional prefactor of $1/s_2$ compared to Eq.~\eqref{eq:Max_t}, as it remains independent of $t_2$.} This results in a compact formula for the Laplace transform of the covariance of the number of records at two distinct times, 
	\begin{align}
		\mathcal{L} \lbrace \text{Cov}\left[M(t_1),M(t_2) \right] \rbrace %\nonumber \\
		%&\equiv \mathcal{L} \lbrace \mathbb{E}(M(t_1)M(t_2))- \mathbb{E}(M(t_1))\mathbb{E}(M(t_2))\rbrace \nonumber \\
		=\frac{w(s_1)+w(s_2)-w(s_1+s_2)}{s_1 s_2 w(s_1)w(s_2)w(s_1+s_2)}
	\end{align}
	in the limit of large times and number of records.
	
	Equation~\eqref{eq:M1_M2_cum} extends the findings of \cite{Benichou:2016a} to arbitrary Markovian processes with nearest-neighbor jumps. In particular, it allows the computation of various two-time functionals for the process. An illustrative example is provided by the conditional survival probability of a Rosenstock trapping problem \cite{Rosenstock:1970}, where traps, present in concentration $q$, are exclusively located on the positive axis \cite{Regnier:2022}. The conditional survival probability $S(t_2|t_1)$, defined as the probability of survival up to time $t_2$ given that the random walk  survived up to time $t_1$, is given by the probability that there is no trap in the newly visited domain of extension $M(t_2)-M(t_1)$. This implies that   $S(t_2|t_1)=\mathbb{E}\left( (1-q)^{M(t_2)-M(t_1)}\right)$. This quantity is fully determined by the knowledge of the two-time distribution Eq.~\eqref{eq:M1_M2_cum}.
	
	\subsection{The $n$-time distributions}
	The previous calculations can be extended to determine $n$-time distributions. For the sake of simplicity, we focus here on the case $0=m_0<m_1 \ldots <m_n$, where at least one new record occurred at each observation time.
	
	First, for the joint distribution of having at least $m_k$ records at times $t_k$ for $k=1$ to $n$, we get in the Laplace domain :
	\begin{align}
		&\mathcal{L} \lbrace \mathbb{P}\left( M(t_1)\geq m_1,  \ldots, \;  M(t_n)\geq m_n \right) \rbrace \nonumber\\
		&=\mathbb{E}\left( \prod_{k=1}^n \int_0^\infty {\rm d}t_k H(t_k-T_{m_k}) e^{-s_k t_k}\right) \nonumber \\
		&=\frac{1}{s_1\ldots s_n} \mathbb{E} \left(\prod_{k=1}^n e^{-s_kT_{m_k}} \right) \nonumber \\
		&=\frac{\mathbb{E} \left(e^{-(s_1+\ldots+s_n)T_{m_1}} \right)}{s_1\ldots s_n}  \mathbb{E}\left( \prod_{k=2}^n e^{-s_k(T_{m_k}-T_{m_1})} \right) \nonumber  \\
		&=\frac{1}{s_1\ldots s_n}e^{-w(s_1+\ldots+s_n)m_1} \mathbb{E}\left( \prod_{k=2}^n e^{-s_kT_{m_k-m_1}} \right) \nonumber  \\
		&=\frac{1}{s_1\ldots s_n} \prod_{k=1}^n e^{-w(\sum_{j=k}^n s_j)(m_{k}-m_{k-1})}\; . \label{eq:n_times_Records}
	\end{align}
	\blue{We recall that the result is exact for the maximum, but asymptotic for describing the records, such that for every $k$, the time scales $s_k$ approach zero (with their ratio remaining finite), and the differences $m_k - m_{k-1}$ behave inversely proportional to $w(s_k)$}. \\

	Then, for the joint distribution of the successive times $T_{M(t_k)}$ at which the records \blue{were} last reached and their associated number of records: 
	\begin{align}
		&\mathcal{L} \lbrace \mathbb{P}\left( M(t_k)= m_k,\; T_{m_k}=y_k; 1\leq k\leq n \right) \rbrace \nonumber  \\
		=&\mathbb{E}\Big( \prod_{k=1}^n \iint_0^\infty {\rm d}t_k {\rm d}y_k H(t_k-T_{m_k})H(T_{m_k+1}-t_k) \nonumber \\
		&\times\delta\left(y_k-T_{m_k} \right) e^{-u_k y_k-s_kt_k}\Big) \nonumber \\
		=&\mathbb{E}\left( \prod_{k=1}^n e^{-(u_k+s_k) T_{m_k}}\frac{1-e^{-s_k\tau_{m_k}}}{s_k}\right) \nonumber
	\end{align}
	\begin{align}
		=&\mathbb{E}\left( e^{-T_{m_1}\sum_{j=1}^n(u_j+s_j)} e^{-\tau_{m_1}\sum_{j=2}^n(u_j+s_j)} \frac{1-e^{-s_1\tau_{m_1}}}{s_1}\right) \nonumber \\
		&\times \mathbb{E}\left( \prod_{k=2}^n e^{-(u_k+s_k) (T_{m_k}-T_{m_1}-\tau_{m_1})}\frac{1-e^{-s_k\tau_{m_k}}}{s_k}\right) \nonumber \\
		=& \frac{1}{s_1\ldots s_n} \prod_{k=1}^n  \Bigg[ e^{-w\left( \sum_{j=k}^n (u_j+s_j) \right)(m_k-m_{k-1})} \nonumber \\
		&\left(w\left(s_k+\sum_{j=k+1}^n(u_j+s_j)\right) -w\left(\sum_{j=k+1}^n (u_j+s_j) \right) \right)  \Bigg].
		\label{eq:n_times_RecordsTimes}
	\end{align}
	
	\begin{figure}[t!]
		\includegraphics[width=0.9\columnwidth]{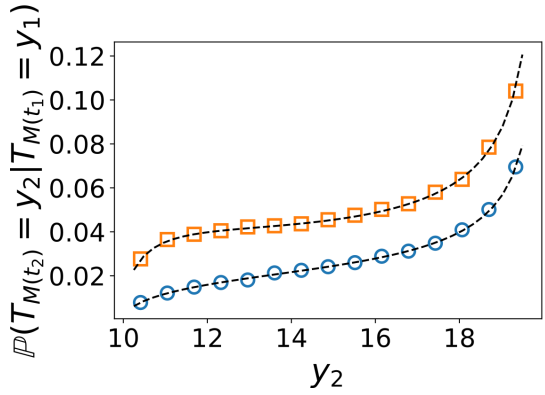}
		\caption{{\bf Conditional last record time.} Distribution of $T_{M(t_2)}$  conditioned on the time $T_{M(t_1)}=2$ (blue circles) or $8$ (orange squares), for the symmetric BM of parameters $D=1$, $t_1=10$, $t_2=20$. The black dashed line corresponds to Eq.~\eqref{eq:Generalised_arcsine}.}
	\end{figure}
	
	Similarly, for the $n$ consecutive backward record times $B_{t_k}$ and their associated number of records: 
	\begin{align}
		&\mathcal{L} \lbrace \mathbb{P}\left( M(t_k)= m_k,\; B_{t_k}=y_k; 1\leq k\leq n \right) \rbrace \nonumber  \\
		=&\mathbb{E}\Big( \prod_{k=1}^n \iint_0^\infty {\rm d}t_k {\rm d}y_k H(t_k-T_{m_k})H(T_{m_k+1}-t_k) \nonumber \\
		&\times\delta\left(t_k-y_k-T_{m_k} \right) e^{-u_k y_k-s_kt_k}\Big) \nonumber \\
		&=\mathbb{E}\left( \prod_{k=1}^n e^{-s_k T_{m_k}}\frac{1-e^{-(s_k+u_k)\tau_{m_k}}}{s_k+u_k}\right) \nonumber     \\
		&= \prod_{k=1}^n  \Bigg[ e^{-w\left( \sum_{j=k}^n s_j \right)(m_k-m_{k-1})} \nonumber \\
		&\frac{w\left(u_k+\sum_{j=k}^ns_j\right) -w\left(\sum_{j=k+1}^n s_j \right) }{s_k+u_k}  \Bigg] \; .
		\label{eq:n_times_Backward}
	\end{align}
	
	Finally, for the $n$ consecutive forward record times $E_{t_k}$ and their associated number of records:
	\begin{align}
		&  \mathcal{L} \lbrace \mathbb{P}\left( M(t_k)= m_k,\; E_{t_k}=y_k; 1\leq k\leq n \right) \nonumber \\
		=&\mathbb{E}\Bigg( \prod_{k=1}^n \iint_0^\infty {\rm d}t_k {\rm d}y_k H(t_k-T_{m_k}) \nonumber \\
		&\times \delta(y_k+t_k-T_{m_{k}+1}) e^{-u_k y_k-s_kt_k}\Bigg) \nonumber \\
		=&\mathbb{E}\left( \prod_{k=1}^n e^{-s_kT_{m_k}}\frac{e^{-u_k \tau_{m_{k}}}-e^{-s_k\tau_{m_k}}}{s_k-u_k}\right) \nonumber \\
		=&\mathbb{E}\left(  e^{-(s_1+\ldots+s_n)T_{m_1}-(s_2+\ldots+s_n)\tau_{m_1}}\frac{e^{-u_1 \tau_{m_{1}}}-e^{-s_1\tau_{m_1}}}{s_1-u_1}\right) \nonumber \\
		&\times\mathbb{E}\left( \prod_{k=2}^n e^{-s_k(T_{m_k}-T_{m_1}-\tau_{m_1})}\frac{e^{-u_k \tau_{m_{k}}}-e^{-s_k\tau_{m_k}}}{s_k-u_k}\right) \nonumber \\
		=&\prod_{k=1}^n \Bigg[ e^{-w\left(\sum_{j=k}^n s_j \right)(m_k-m_{k-1})}\nonumber  \\
		&\frac{w\left(\sum_{j=k}^n s_j \right)-w\left(u_k+\sum_{j=k+1}^n s_j \right)}{s_k-u_k} \Bigg]. \label{eq:n_times_Forward}
	\end{align}
	Note that we do not derive the formula when two successive $m_k$ are equal: as a matter of fact, one can derive these values from the one with distinct successive $m_k$ as for Eq.\; \eqref{eq:Equality_mk}.
	The multi-time joint distribution of $M(t)$, $E_t$ and $B_t$ can similarly be obtained. It is explicitly given in Appendix \ref{app:MEB}. 
	
	As an explicit example of multi-time distribution, one can compute the joint distribution of the time at which the last record event occurred at two different times $t_1$ and $t_2$. This can be done using Eq.~\eqref{eq:n_times_RecordsTimes} where $k=2$. We integrate over $m_1$ and $m_2$ and consider the times $y_1<t_1<y_2<t_2$, such that for the BM:
	\begin{align}
		\mathbb{P}\left( T_{M(t_2)}=y_2|T_{M(t_1)}=y_1 \right)=\frac{\sqrt{y_2-t_1}}{\pi (y_2-y_1)\sqrt{t_2-y_2}}.
		\label{eq:Generalised_arcsine}
	\end{align}
	This provides a generalization of the celebrated arcsine law \cite{Levy:1940} for the distribution of the time for reaching the maximum in a time interval. In particular, we observe that having a time $y_1$ of observation of the last record in the interval $[0,t_1]$ close to $t_1$ increases the probability of having a large time of observation of the last record up to $t_2$. Thus, there are high correlations between successive last record times that cannot be accounted for by  single-time expressions.\\
	
	\blue{As an other example of application, for any $ n $, we can compute the difference between the distribution of record events and that of independent record occurrences, akin to the methodology outlined in \cite{Regnier:2022}. This investigation holds particular relevance when evaluating the probability of a maximum or record trajectory sampled at times $ t_1, t_2, \ldots, t_n $ to discern properties of the RW, as explored in \cite{Benichou:2016b}. In the scenario where the time intervals between samplings are significantly large, $ t_1\ll t_2 \ll \ldots \ll t_n $, or equivalently $ s_1\gg s_2\gg \ldots \gg s_n $ with $ m_kw(s_k) $ held constant (considering only typical events), we have:
		\begin{align} 
			&\mathcal{L} \lbrace \mathbb{P}\left( M(t_k)\geq m_k; 1\leq k\leq n \right) \rbrace- \mathcal{L}\left\lbrace\prod_{k=1}^n  \mathbb{P}\left( M(t_k)\geq m_k \right) \right\rbrace \nonumber \\
			&\sim  \mathcal{L}\left\lbrace\prod_{k=1}^n  \mathbb{P}\left( M(t_k)\geq m_k \right) \right\rbrace  \nonumber \\
			&\times\Bigg[-\sum_{k=1}^n w'(s_k)s_{k+1}m_k +w(s_k)m_{k-1} \Bigg] \; 
		\end{align}
		Utilizing the asymptotics:
		\begin{align}
			w'(s_k)s_{k+1}m_k \propto \frac{s_{k+1}}{s_k}, \; w(s_k)m_{k-1} \propto \frac{w(s_k)} {w(s_{k-1})} \; ,
		\end{align}
		we deduce that the process decorrelates when there are large time intervals between samplings for processes where $w(s)$ approaches $ 0 $ as $ s $ tends to $ 0 $ \footnote{\blue{Note that this condition is not always met for aBM and aRTP, so that correlations are not decaying in time.}}. In such instances, the decorrelation follows an algebraic pattern with an exponent of $ 1/2 $ for the symmetric case and $ 1 $ for the biased case. The rate of convergence is determined by the largest of the ratios $ s_k/s_{k-1} $, or equivalently $ t_{k-1}/t_k $. In essence, employing the distribution of maxima at $ n $ discrete times under the assumption of independence among maxima incurs an error that scales proportionally with the smallest ratio of sampling times, with the exponent contingent upon the symmetry properties of the walk.
	} \\
	
	We stress that, by determining these multiple time distributions, we have  provided a {\it complete} characterisation of the record process.

	\section{Generalisation to non-identical renewal times} 
	
	One can easily generalise the preceding results to the case of  non identically distributed but still independent renewal times. An important explicit example involving aging renewal times is provided by the resetting RW \cite{Evans:2011,Evans:2018,Kumar:2023,Singh:2021,Evans:2020} (see Appendix~\ref{app:reflecting} for the other \blue{important example} of a RW with a reflecting boundary condition at initial position).  
	
	In the resetting RW, at every time step, the RW gets back to its initial position with probability $1-\lambda=1-e^{-r}$ ($r$ is the resetting rate). These types of processes have attracted a lot of attention in the past few years, and are well documented. In particular, the record age distribution with resetting (where we reset both the position and velocity to the initial value) knowing that $m$ records have been achieved, $\hat{F}_m^r(s)$, is given by (see Appendix~\ref{app:reset} for details):
	\begin{align}
		1-\hat{F}_{m-1}^r(s)&=s\frac{1-\hat{F}(s+r)}{s+r\hat{F}(s+r)^m} \nonumber \\
		&=s\frac{w(s+r)}{s+re^{-w(s+r)m}} \equiv g(s,m) \blue{,} \label{eq:record_age_reset}
	\end{align}
	where $\hat F$ and $w$ stand for the corresponding quantities without resetting\blue{.}
	
	Following the methodology developed in the previous sections, in particular following the steps described in Eqs.~\eqref{eq:n_times_Records} to \eqref{eq:n_times_Forward}, we get for the $n$ records distributions, 
	\begin{align}
		&\mathcal{L} \lbrace \mathbb{P}\left( M(t_1)\geq m_1, \ldots  \; M(t_n)\geq m_n \right) \rbrace \nonumber \\
		&=\frac{1}{s_1\ldots s_n} \mathbb{E} \left(e^{-(s_1+\ldots+s_n)T_{m_1}} \right) \mathbb{E}\left( \prod_{k=2}^n e^{-s_k(T_{m_k}-T_{m_1})} \right) \nonumber  \\
		&=\frac{1}{s_1\ldots s_n}e^{-\int_0^{m_1} g(s_1+\ldots+s_n,m) {\rm d}m} \nonumber \\
		&\times \mathbb{E}\left( e^{-s_2(T_{m_2}-T_{m_1})}\ldots e^{-s_n(T_{m_n}-T_{m_1})} \right) \nonumber\\
		&=\prod_{k=1}^n \left( \frac{e^{-\int_{m_{k-1}}^{m_{k}} g\left(\sum_{j=k}^n s_j,m \right) {\rm d}m }}{s_k} \right)\; . \label{eq:Records_reset_multiple_times}
	\end{align}
	
	Then, for the successive times at which the records were last reached and their current number: 
	\begin{align}
		&\mathcal{L} \lbrace \mathbb{P}\left( M(t_k)= m_k,\; T_{m_k}=y_k; \;1 \leq k \leq n \right) \rbrace \nonumber \\
		&= \prod_{k=1}^n \Bigg[ e^{-\int_{m_{k-1}}^{m_k} {\rm d}m g\left( \sum_{j=k}^n (u_j+s_j),m \right)}\nonumber \\
		&\times \frac{g\left(s_k+\sum_{j=k+1}^n(u_j+s_j),m_k\right)-g\left(\sum_{j=k+1}^n (u_j+s_j),m_k \right) }{s_k}\Bigg]\label{eq:n_records_time_reset}
	\end{align}
	
	Similarly, for the $n$ consecutive backward record times and their associated number of records: 
	\begin{align}
		&\mathcal{L} \lbrace \mathbb{P}\left( M(t_k)= m_k,\; B_{t_k}=y_k; 1\leq k\leq n \right) \rbrace \nonumber \\
		&= \prod_{k=1}^n \Bigg[ e^{-\int_{m_{k-1}}^{m_k} {\rm d}m g\left( \sum_{j=k}^n s_j,m \right)} \nonumber \\
		& \times \frac{g\left(u_k+\sum_{j=k}^n s_j,m_k\right)-g\left(\sum_{j=k+1}^n s_j,m_k \right)}{s_k+u_k}  \Bigg]
		\label{eq:Backward_record_reset}
	\end{align}
	
	Finally, for the $n$ consecutive forward record times and their associated number of records:
	\begin{align}
		&\mathcal{L} \lbrace \mathbb{P}\left( M(t_k)= m_k,\; E_{t_k}=y_k; 1\leq k\leq n \right) \nonumber \\
		&=\prod_{k=1}^n \Bigg[ e^{-\int_{m_{k-1}}^{m_k} g\left(\sum_{j=k}^n s_j,m \right){\rm d}m } \nonumber \\
		&\times\frac{g\left(s_k+\sum_{j=k+1}^n s_j,m_k \right)-g\left(u_k+\sum_{j=k+1}^n s_j,m_k \right)}{s_k-u_k} \Bigg]
		\label{eq:Forward_record_reset}
	\end{align}
	
\begin{figure}[t!]
		\centering
		\includegraphics[width=\columnwidth]{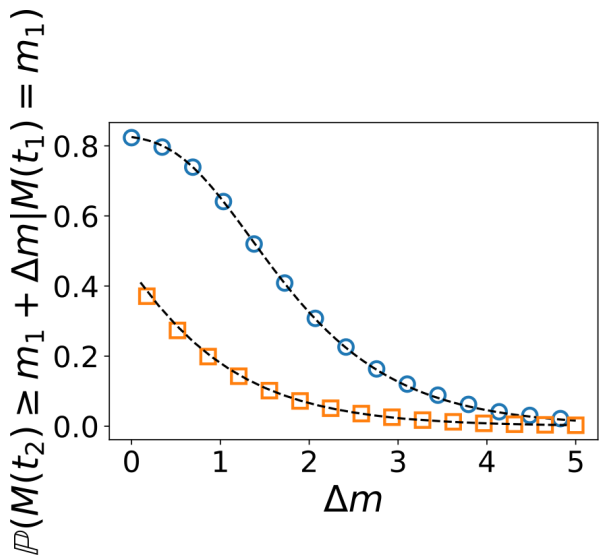}
		\caption{Conditional tail distribution of $M(t_2)$ knowing that $M(t_1)=1$ (blue circles) $=3$ (orange \blue{squares}) at time $t_1=10$ and $t_2=20$ for the resetting BM of parameters $r=1$, $D=1$. Black dashed lines represent the result of numerical Laplace inversion of Eq.~\eqref{eq:Records_reset_multiple_times} for $n=2$.}
		\label{fig:Mt1t2_reset}
	\end{figure}

	As an example, we provide in Fig.~\ref{fig:Mt1t2_reset} the conditional distribution of the number of records $M(t_2)$ at time $t_2$ knowing the number of records $M(t_1)$ at time $t_1$ for different values of $M(t_1)$, for the resetting BM. Numerically Laplace inverting Eq.~\eqref{eq:Records_reset_multiple_times} for the resetting BM, we observe that having a large number of records at early time decreases the probability to observe new records at later times. Indeed, if the RW reaches a large number of records (and thus a position far from its origin) before resetting, it will become harder for later trajectories after reset to overcome a similar number of records.

\begin{figure}[t!]
		\centering
		\includegraphics[width=\columnwidth]{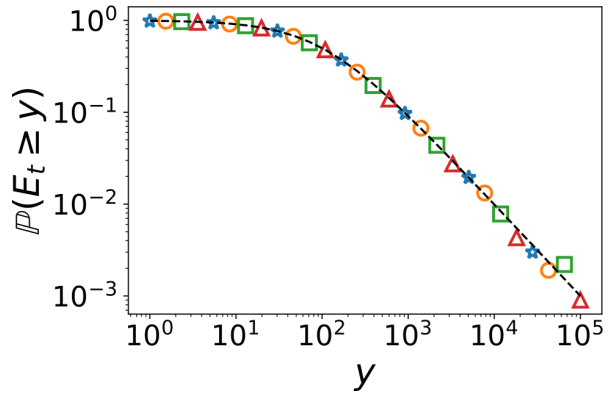}
		\caption{Time $E_t$ until the next record breaking event from time $t=100$, for BM (blue stars), aBM (orange circles), RTP (green squares) and aRTP (red triangles) of resetting rate $r=1$, diffusion coefficient $D=1$, $T=1$, $v=1$, $\mu=0.5$. The asymptotic expression of Eq.~\eqref{eq:Forward_universal} is in black dashed line. }
		\label{fig:Et_reset}
	\end{figure}

	Additionally, from Eqs.~\eqref{eq:Backward_record_reset} and \eqref{eq:Forward_record_reset}, one can derive the asymptotic distributions of $B_t$ and $E_t$. We note that Eq.~\eqref{eq:n_records_time_reset} gives the same formula as obtained in \cite{Singh:2021} for the single time distribution and thus $B_t$ is asymptotically uniformly distributed in the interval $[0,t]$ at large times $t$, independently of the process. In fact, one can find a similar result for the forward record time $E_t$. In the limit $t,y\gg 1/r$ corresponding to $s,u\ll r$, we have a universal form for the distribution of $E_t$:
	\begin{align}
		\mathcal{L}\lbrace \mathbb{P}(E_t=y) \rbrace  &\sim  \int_0^\infty {\rm d}m \frac{r}{r+se^{-w(r)m}} \frac{w(r)}{s-u} \nonumber \\
		& \times \left(\frac{s}{s+re^{w(r)m}}-\frac{u}{u+re^{w(r)m}} \right) \nonumber \\
		&=\int_0^1 {\rm d}M \frac{r}{(r+sM)^2}  \frac{1 }{uM+r }
	\end{align}
	and by Laplace inversion, 
	\begin{align}
		\mathbb{P}(E_t \geq y)  &=\int_0^1 {\rm d}M rt e^{-Mrt}   e^{-M r y} \sim \frac{t}{t+y} \label{eq:Forward_universal}
	\end{align}
	This is exactly the result one expects for the time to break the next record from a series of i.i.d. random variables at discrete times $\left(X_{t'} \right)_{t'\geq 0}$ of cumulative $F(x)=\mathbb{P}(X\leq x)$:
	\begin{align}
		&\mathbb{P}(E_t\geq y) \nonumber \\
		&=\int_0^\infty {\rm d} m \mathbb{P}\left(X_{t+k}\leq m;1 \leq k\leq y \right)\mathbb{P}\left(\underset{t'\leq t}{\max}(X_{t'})=m \right) \nonumber \\
		&=\int_0^\infty {\rm d} m t F(m)^{t+y-1} \frac{{\rm d}}{{\rm d}m} F(m) =\frac{t}{t+y} \; .
	\end{align}
	Similarly to what was shown in \cite{Singh:2021} in the case of the backward time, this universal result holds for general stochastic process (arbitrary $w(s)$) as it stems from the independence between reset trajectories, as it is shown in Fig.~\ref{fig:Et_reset}.

	\section{Summary of the results} 
	In summary, we have developed a general approach based on renewal theory to derive multiple-time statistics for the number of records in one dimension for nearest-neighbor Markovian Random Walks, along with various observables associated with the record process. The formulae obtained are compact in the Laplace domain and can be easily used to gain insights into the statistics of records. These multi-time distributions provide a complete characterisation of the statistics of records. We have demonstrated that this method can be generalized to resetting random walks, thereby determining the complete statistics of record observables.
	
	\appendix
	
	\renewcommand{\thefigure}{Figure S\arabic{figure}}
	\renewcommand{\figurename}{}
	\setcounter{figure}{0}
	
	\title{Records statistics of 1d Markovian Processes: Appendix}
	
	\section{First-Passage Time distributions of aBM and aRTP}
	\label{app:FPT}
	
	The First-Passage Time (FPT) distribution is the key ingredient for that of the number of records, see Eq.~\eqref{eq:Max_t}.
	For self-consistency, we provide in this Appendix the standard computations  \cite{Redner:2001,Hughes:1995,Klinger:2022} of the FPT distributions of both the asymmetric Brownian Motion (aBM) and the asymmetric Run-and-Tumble Particle (aRTP), in discrete and continuous time. 
	
	\subsection{Asymmetric Brownian Motion (aBM)}
	\subsubsection{Discrete time} \label{sec:discrete_aBM}
	In the discrete asymmetric nearest-neighbour RW model (or discrete aBM) on the $1d$ lattice, the random walker (RW) hops to the site on its right with probability $p$ and the one on its left with probability $1-p$. 
	
	To obtain the FPT of the RW to position $m$, we consider the backward master equation in discrete time, by noting  $P(x,t)$ the probability to reach position $m$ for the first time starting at position $x<m$,
	\begin{align}
		P(x,t+1)=pP(x+1,t)+(1-p) P(x-1,t) \label{eq:Back_BM_discrete}
	\end{align}
	with boundary condition $P(m,t)=\delta(t)$ and $P(-\infty,t)=0$. The record age distribution is given by $P(0,t)$ for $m=1$. By discrete Laplace transform of Eq.~\eqref{eq:Back_BM_discrete}, 
	\begin{align}
		\mathcal{L}\lbrace f(t) \rbrace=\hat{f}(s)=\sum_{t=0}^\infty e^{-st} f(t) \; ,
	\end{align}
	we obtain 
	\begin{align}
		e^{s}\hat{P}(x,s)=p\hat{P}(x+1,s)+(1-p)\hat{P}(x-1,s) \; . 
	\end{align}
	From the boundary conditions, we deduce that
	\begin{align}
		\hat{P}(x,s)=\left[ \frac{e^{s}}{2(1-p)}\left(1-\sqrt{1-4p(1-p)e^{-2s}} \right)\right]^{m-x} \label{eq:F_aBM_discrete}
	\end{align}
	For $m=1$ and $x=0$, we deduce the record age distribution for the discrete aBM,
	\begin{align}
		\hat{F}(s)=\frac{e^{s}}{2(1-p)}\left(1-\sqrt{1-4p(1-p)e^{-2s}} \right) \label{eq:F_record_aBM_discrete} \; .
	\end{align}\\
	
	\vspace{-1cm}
	\subsubsection{Continuous time} \label{sec:continuous_aBM}
	In the aBM model on the $1d$ line, the time evolution of the position $X_t$ of the RW is given by the Langevin equation with constant external force $\mu$ (the bias), 
	\begin{align}
		\frac{{\rm d}X_t}{{\rm d}t}=\mu+\eta_t
	\end{align}
	where the stochastic process $\left( \eta_t\right)_{t\geq 0}$ is a Gaussian white noise, 
	\begin{align}
		\left\langle \eta_t \right\rangle= 0  , \; \left\langle \eta_t \eta_{t'} \right\rangle= 2D\delta(t-t') \; .
	\end{align}
	To obtain the FPT to position $m$ of the RW, we consider the backward master equation in continuous time, 
	\begin{align}
		\partial_t p(x,t)=\left( D\partial^2_{x,x}+\mu\partial_x \right)p(x,t) \label{eq:Back_BM}
	\end{align}
	with boundary condition $p(m,t)=\delta(t)$ and $p(-\infty,t)=0$ such that $p(x,t)$ stands for the probability that the RW initially at position $x$ arrived at position $m>0$ for the first time at time $t$. The record age distribution is given by $p(0,t)$ for $m=1$.\\
	To solve Eq.~\eqref{eq:Back_BM}, we make a Laplace transformation. The Laplace transform of a function $f(t)$ is defined in the continuum setting as 
	\begin{align}
		\mathcal{L}\lbrace f(t) \rbrace=\hat{f}(s)\equiv \int_0^\infty{\rm d}t e^{-st} f(t) 
	\end{align}
	Using that $p(x<m,t=0)=0$, Eq.~\eqref{eq:Back_BM} becomes, once Laplace transformed (in time), 
	\begin{align}
		\left(D\partial^2_{x,x}+\mu\partial_x -s\right)\hat{p}(x,s)=0 \label{eq:FPT_eq_BM}
	\end{align}
	and the initial condition $\hat{p}(x=m,s)=1$. We search an exponential solution of the form $\hat{p}(x,s)=e^{-w(s)(m-x)}$ where $w(s)$ is positive ($\hat{p}(-\infty,s)=0$). This results in  the following equation for $w(s)$, 
	\begin{align}
		Dw(s)^2+\mu w(s)-s=0
	\end{align}
	which positive solution is
	\begin{align}
		w(s)=-\frac{\mu}{2D}+\frac{\sqrt{\mu^2+4Ds}}{2D} \; .
	\end{align}
	Finally, we have the Laplace transform of first passage time at $m$ starting at position $x$, 
	\begin{align}
		\hat{F}(s,x)\equiv \hat{p}(x,s)=e^{-w(s)(m-x)} \; . \label{eq:F_aBM}
	\end{align}
	which provides the record age distribution for $m=1$ \blue{(we set a scale needed to traverse in order to make a new record at $1$, see Appendix~\ref{App:discr_cont} for details)}
	%\blue{(the reason for taking $m=1$ will be explained in more details in the next section; the idea is to make the correspondence with the discrete setting)} 
	and $x=0$,
	\begin{align}
		\hat{F}(s)=e^{-w(s)} \; . \label{eq:F_record_aBM}
	\end{align}
	Of note, the discrete model gives asymptotically the continuous model for $p=\frac{1+\mu}{2}$.
	
	\onecolumngrid
	
	\subsection{Asymmetric Run-and-Tumble Particle (aRTP)}
	\subsubsection{Discrete time} \label{sec:discrete_aRTP}
	In the discrete asymmetric Run-and-Tumble Particle (aRTP) model on the $1d$ lattice, the RW has two states, the positive state in which it moves to the neighbouring site on its right, and a negative state in which it moves to the neighbouring site on its left. At each time step, the RW can change its state: it goes from the positive to negative state with probability $p_+$ and from the negative to positive state with probability $p_-$.\\
	To obtain the FPT of the RW to position $m$, we start from the backward master equation in discrete time,
	\begin{align}\label{eq:Back_RTP_discrete}
		\begin{cases}
			P_+(x,t+1)= p_+ P_+(x+1,t)+(1-p_+)P_-(x-1,t) \\ 
			P_-(x,t+1)= p_- P_-(x-1,t)+ (1-p_-)P_+(x+1,t)
		\end{cases}
	\end{align}
	with boundary condition $P_+(m,t)=\delta(t)$ and $P_+(-\infty,t)=P_-(-\infty,t)=0$ such that $P_+(x,t)$ ($P_-(x,t)$) stands for the probability that the RW initially at position $x$ going to the right  (resp. going to the left) arrived at position $m>0$ for the first time at time $t$ (one can only arrive at position $m>x$ by doing a last step to the right). The record age distribution is given by $P_+(0,t)$ for $m=1$. We discrete Laplace transform both equations to obtain
	\begin{align}\label{eq:Back_RTP_discrete}
		\begin{cases}
			e^s\hat{P}_+(x,s)=p_+ \hat{P}_+(x+1,s)+(1-p_+)\hat{P}_-(x-1,s) \\ 
			e^s \hat{P}_-(x,s)= p_- \hat{P}_-(x-1,s)+ (1-p_-)\hat{P}_+(x+1,s)
		\end{cases}
	\end{align}
	Then, by expressing $\hat{P}_-$ as a function of $\hat{P}_+$ using the first line of Eq.~\eqref{eq:Back_RTP_discrete} and replacing in the second line, we have an equation solely on $\hat{P}_+$,
	\begin{align}
		\begin{cases}
			\hat{P}_-(x,s)=\frac{e^s\hat{P}_+(x+1,s)-p_+ \hat{P}_+(x+2,s)}{1-p_+} \\
			p_+ e^s \hat{P}_+(x+2,s)-(e^{2s}-1+(p_+ +p_-))\hat{P}_+(x+1,s)+p_- e^s\hat{P}_+(x,s)=0 \label{eq:FPT_eq_RTP_discrete}
		\end{cases}
	\end{align}
	This results in the following expression for $\hat{P}_+(x,s)$,
	\begin{align}
		\hat{P}_+(x,s)=\left[ \frac{1 + (-1 + p_+ + 
			p_-) e^{-2s} - \sqrt{-4 p_- p_+ e^{-2s} + (1 + (-1 + p_- + p_+) e^{-2s})^2}}{2p_- e^{-s}} \right]^{m-x}
	\end{align}
	and for $x=m-1$, we deduce the record age distribution for the discrete aRTP,
	\begin{align}
		\hat{F}(s)=\frac{1 + (-1 + p_+ + 
			p_-) e^{-2s} - \sqrt{-4 p_- p_+ e^{-2s} + (1 + (-1 + p_- + p_+) e^{-2s})^2}}{2p_- e^{-s}}  \label{eq:F_aRTP_discrete} \; .
	\end{align}
	
	\subsubsection{Continuous time} \label{sec:continuous_aRTP}
	In the asymmetric Run-and-Tumble Particle (aRTP) model, the RW has two states, the positive state in which it moves at speed $v_+$ towards the right side of the interval, and a negative state in which it moves at speed $v_-$ towards the left side of the interval. The RW goes from the positive state to the negative state at time rate $1/2T_+$ and from the negative to positive state at rate $1/2T_-$. \blue{These} tumbling events occur independently.\\
	In other words, the time evolution of the position $X_t$ of the RW is given by (we note $\mu$ the bias and $v$ the average absolute speed),
	\begin{align}
		\frac{{\rm d}X_t}{{\rm d}t}=\frac{v_+-v_-+\sigma(t)(v_++v_-)}{2}=\mu+\sigma(t)v
	\end{align} 
	where $\left( \sigma(t) \right)_{t\geq 0}$ is a Poisson process which switches between values $-1$ and $1$ with rates $1/2T_+$ (for $+1$ to $-1$) and $1/2T_-$ (for $-1$ to $+1$), such that at any time $t$ and for infinitesimal time step ${\rm d}t$,
	\begin{align}
		\sigma(t+{\rm d}t)= 
		\begin{cases}
			\sigma(t) \text{ with probability } 1-\frac{{\rm d}t}{2T_{\sigma (t)}}\\
			-\sigma(t) \text{ with probability } \frac{{\rm d}t}{2T_{\sigma (t)}}
		\end{cases}
	\end{align}
	To obtain the FPT of this model to position $m$, we consider the two coupled backward master equation, 
	\begin{align}\label{eq:Back_RTP}
		\begin{cases}
			\partial_t p_+(x,t)=v_+\partial_xp_+(x,t)+\frac{1}{2T_+}(p_-(x,t)-p_+(x,t)) \\ 
			\partial_t p_-(x,t)=-v_-\partial_xp_-(x,t)+\frac{1}{2T_-}(p_+(x,t)-p_-(x,t))
		\end{cases}
	\end{align}
	with boundary condition $p_+(m,t)=\delta(t)$ such that $p_+(x,t)$ ($p_-(x,t)$) stands for the probability that the RW initially at position $x$ with speed $v_+$ (resp. $-v_-$) arrived at position $m>0$ for the first time at time $t$ (one can only arrive at position $m$ with positive speed). The record age distribution is given by $p_+(0,t)$ for $m=1$. We Laplace transform both equations to obtain
	\begin{align}\label{eq:Back_RTP_Lap}
		\begin{cases}
			s\hat{p}_+(x,s)=v_+\partial_x\hat{p}_+(x,s)+\frac{1}{2T_+}(\hat{p}_-(x,s)-
			\hat{p}_+(x,s)) \\ 
			s\hat{p}_-(x,s)=-v_-\partial_x \hat{p}_-(x,s)+\frac{1}{2T_-} (\hat{p}_+(x,s)-\hat{p}_-(x,s))
		\end{cases}
	\end{align}
	Then, by expressing $\hat{p}_-(x,s)$ as a function of $\hat{p}_+(x,s)$ and its derivatives using the first line of Eq.~\eqref{eq:Back_RTP_Lap} and replacing in the second line, we have an equation solely on $\hat{p}_+(x,s)$,
	\begin{align}
		\begin{cases}
			\hat{p}_-(x,s)= \left(1+2sT_+\right) \hat{p}_+(x,s)-2v_+T_+\partial_x \hat{p}_+(x,s)\\
			\left(s (T_- + T_+ + 2 s T_- T_+)+\left( T_- (1 + 2 s T_+) v_- - (1 + 2 s T_-) T_+ v_+ \right)\partial_x    -2 T_- T_+ v_- v_+  \partial_{x,x}^2\right)\hat{p}_+(x,s)=0 \label{eq:FPT_eq_RTP}
		\end{cases}
	\end{align}
	We have the boundary condition $\hat{p}_+(m,s)=1$ and $\hat{p}_{\pm}(-\infty,s)=0$. This results in the following formula for $w(s)$,
	\begin{align}
		w(s)=\frac{ 2 s T_+T_- (v_--v_+)+T_-v_--T_+ v_++\sqrt{(T_+ v_+-T_-v_--2 s T_+T_- ( v_--
				v_+))^2+8 s T_- T_+ v_- v_+ (2 s T_- T_++T_-+T_+)}}{4 T_- T_+ v_-
			v_+}
	\end{align}
	Because we are interested only in the case where the particles starts with positive velocity, we have the first passage at position $m$ starting at position $x$ with positive speed,
	\begin{align}
		\hat{F}(x,s)=\hat{p}_+(x,s)=e^{-w(s)(m-x)} \; . \label{eq:F_aRTP}
	\end{align}
	and the record age distribution in the Laplace domain is obtained taking $m=1$
	\blue{(similarly to the aBM, see Eq.~\eqref{eq:F_aBM})} 
	%\blue{(similarly to the aBM case, the reason for taking $m=1$ will be explained in more details in the next section)} 
	and $x=0$,
	\begin{align}
		\hat{F}(s)=e^{-w(s)} \; . \label{eq:F_record_aRTP}
	\end{align}
	Of note, the discrete model gives asymptotically the continuous model for $v_+=v_-=1$, $p_+=1-1/2T_+$ and $p_-=1-1/2T_-$. In the main text, we present the model with tumbling rate $T=T_+=T_-$ and with external constant force $f=\mu$ such that the speed in each direction is $v_+=v+\mu$ and $v_-=v-\mu$, as it is not described by the discrete aRTP model and has attracted more attention from the physics community \cite{LeDoussal:2020,DeBruyne:2021,Cinque:2021}.

	\subsection{Real time expressions} \label{app:C_expressions}
	We first recall the real-time expressions of the time to reach level $m$ for the first time starting at position $x=0$ (see \eqref{eq:F_aBM} and \eqref{eq:F_aRTP}) in the case of the aBM and aRTP. For the aBM \cite{Redner:2001}, 
	\begin{align}
		F(m,t)&=\mathcal{L}^{-1}\left\lbrace e^{-w(s)m }\right\rbrace_{s \to t}\\
		&=\frac{1}{2\pi i} \int_{-i \infty}^{i \infty} {\rm d s} e^{st } e^{-w(s)m } \\
		&=\frac{m}{\sqrt{4\pi D t^3}}\exp\left[-\frac{(m-\mu t)^2}{4 D t} \right] \label{eq:FPT_C_aBM}
	\end{align}
	for the aRTP \cite{Lopez:2014,DeBruyne:2021}, 
	\begin{align}
		F(m,t)=e^{-t/2T_+}\delta \left(t-\frac{m}{v_+} \right)+m H\left(t-\frac{m}{v_+} \right)\frac{e^{-\frac{(T_+v_++T_-v_-)t+(T_--T_+)m}{(v_++v_-)2T_+T_-}}}{ \sqrt{4T_+T_-(v_+t-m)(v_-t+m)}}I_1\left(\frac{\sqrt{(v_+t-m)(v_-t+m)}}{(v_++v_-)\sqrt{T_+T_-}} \right) \label{eq:FPT_C_aRTP}
	\end{align}
	From this, we can obtain the inverse Laplace transform of $w(s)/s$, as
	\begin{align}
		g(t)=\mathcal{L}^{-1}\left\lbrace w(s)/s \right\rbrace_{s \to t}&=\mathcal{L}^{-1}\left\lbrace \lim_{m\to 0} (1-e^{-w(s)m})/sm \right\rbrace_{s \to t}=\lim_{m\to 0} \frac{1-\int_0^t F(t',m){\rm d}t'}{m}=\lim_{m\to 0} \frac{\int_t^\infty F(t',m){\rm d}t'}{m}
	\end{align}
	For the aBM, it results in 
	\begin{align}
		g(t)&=\int_t^\infty \frac{{\rm d}t'}{\sqrt{4 \pi D t'^3}}e^{-\mu^2t'/4D}
	\end{align}
	and for the aRTP,
	%\begin{align}
	%    g(t)=\int_t^\infty e^{-t'/2T}I_1\left(\frac{t'\sqrt{v^2-\mu^2}}{2Tv} \right) \frac{{\rm d}t'}{2T t'\sqrt{v^2-\mu^2}}
	%\end{align}
	\begin{align}
		g(t)=\int_t^\infty {\rm d}t'  \frac{e^{-\frac{(T_+v_++T_-v_-)t'}{(v_++v_-)2T_+T_-}}}{ t' \sqrt{4T_+T_-v_+v_-}}I_1\left(\frac{t'\sqrt{v_+v_-}}{(v_++v_-)\sqrt{T_+T_-}} \right) 
	\end{align}
	From this, we derive the inverse Laplace transform of $w(s)$, first for the aBM,
	\begin{align}
		f(t)&=\mathcal{L}^{-1}\left\lbrace w(s) \right\rbrace_{s \to t}=\partial_t g(t)\\
		&=-\frac{1}{\sqrt{4 \pi D t^3}}\exp\left[-\frac{\mu^2 t}{4 D} \right]
	\end{align}
	and for the aRTP,
	\begin{align}
		f(t)=-\frac{e^{-\frac{(T_+v_++T_-v_-)t}{(v_++v_-)2T_+T_-}}}{ t \sqrt{4T_+T_-v_+v_-}}I_1\left(\frac{t\sqrt{v_+v_-}}{(v_++v_-)\sqrt{T_+T_-}} \right).
	\end{align}
	Finally, using the standard properties of double inverse Laplace transforms, we have that (for $y,t>0$, see Eq.~\eqref{eq:1_Backward_record} of the main text)
	\begin{align}
		\mathbb{P}\left(M(t)=m , B_t=y \right)&=\mathcal{L}^{-1}\left\lbrace e^{-w(s)m}\frac{w(s+u)}{s+u}\right\rbrace_{s\to t, u\to y}\\
		&= \int_0^t \int_0^y {\rm d}t' {\rm d}y' \mathcal{L}^{-1}\left\lbrace e^{-w(s)m}\right\rbrace_{s\to t-t', u\to y-y'} \mathcal{L}^{-1}\left\lbrace \frac{w(s+u)}{s+u}\right\rbrace_{s\to t', u\to y'} \\
		&=\int_0^t \int_0^y {\rm d}t' {\rm d}y' F(m,t-t') \delta(y-y') g(t')\delta(t'-y') \\
		&=F(m,t-y)g(y)
	\end{align}
	and for the forward record time (see Eq.~\eqref{eq:forward_record_Laplace}  of the main text), 
	\begin{align}
		\mathbb{P}\left(M(t)=m , E_t=y \right)&=\mathcal{L}^{-1}\left\lbrace e^{-w(s)m}\frac{w(s)-w(u)}{s-u}\right\rbrace_{s\to t, u\to y}\\
		&= \int_0^t \int_0^y {\rm d}t' {\rm d}y' \mathcal{L}^{-1}\left\lbrace e^{-w(s)m}\right\rbrace_{s\to t-t', u\to y-y'} \mathcal{L}^{-1}\left\lbrace \frac{w(s)-w(u)}{s-u}\right\rbrace_{s\to t', u\to y'} 
	\end{align}
	which leads to
	\begin{align}
		\mathbb{P}\left(M(t)=m , E_t=y \right)  
		=-\int_0^t \int_0^y {\rm d}t' {\rm d}y' F(m,t-t') \delta(y-y') f(t'+y')
		=-\int_0^t F(m,t')f(t'+y) {\rm d}t' \; ,
	\end{align}
	corresponding to Eqs.~\eqref{eq:Forward_aBM_t} and \eqref{eq:Forward_aRTP_t} of the main text.
	
	\section{Correspondence between discrete and continuous settings}\label{App:discr_cont}
	
	In this section, we provide details on the passage from the discrete to the continuum descriptions. To this end, we start by discretizing the $1d$ line with spacings $\Delta m$ and we consider directly $X_t$ the continuous process (e.g. aBM or aRTP). 
	\blue{The idea is that we now define a new record as occurring only when the continuous RW exceeds its previous position by $\Delta m$. Consequently, if $k$ records have occurred for the RW, it precisely indicates that it has reached position $k\Delta m$ but has not yet reached $(k+1)\Delta m$. It is immediately evident that by setting $\Delta m=1$, as done in Appendix~\ref{app:FPT}, the number of records and the maximum position coincide (more precisely, the integer part of the maximum, which can be disregarded when the maximum is large).}
	Then, the $\tau_k$ correspond to the times for the continuous process having reached $k \Delta m=m$ to exceed value $(k+1)\Delta m=m+\Delta m$, of density exactly given by $F(\Delta m,t)$ (see Eqs.~\eqref{eq:FPT_C_aBM} and \eqref{eq:FPT_C_aRTP} above). This implies that the (discretized) maximum distribution is indeed given by a renewal formalism but with continuous record ages directly.
	\begin{align}
		\mathbb{P}\left( M(t)\geq m\right)=\mathbb{P}\left( \sum_{k=0}^{m/\Delta m-1} \tau_k \leq t \right),
	\end{align}
	from which using the Laplace transform we get
	\begin{align}
		\mathcal{L}\lbrace \mathbb{P}\left( M(t)\geq m\right) \rbrace &=\frac{1}{s}\prod_{k=0}^{m/\Delta m-1} \hat{F}(\Delta m,s)\\
		&=\frac{1}{s}\exp\left[-\sum_{k=0}^{m/\Delta m-1} w(s)\Delta m \right]\\
		&= \frac{1}{s}\exp\left[- w(s)m \right].
	\end{align}
	Of note, the sum of the $\tau_k$, $\sum_{k=0}^{m/\Delta m} \tau_k=T_m$ is exactly the time to reach position $m$ for the continuous process (no dependence on $\Delta m$ as long as $m$ lies on the lattice of step size $\Delta m$). For the backward record time, the probability that the maximum $M(t)$ of value between $m$ and $m+\Delta m$ (discretization step) occurred at time $t-y=t-B_t$ is given exactly by the probability that  $m$ was reached at time $T_m=t-y$ and the position $m+\Delta m$ was not reached in the time $y$ left (with $y\leq t$). This implies that
	\begin{align}
		&\mathcal{L} \lbrace \mathbb{P}\left( B_t=y,M(t) \in [m,m+\Delta m] \right)\rbrace \nonumber  \\
		&\iint {\rm d}t{\rm d}y e^{-uy-st}\mathbb{E}\left( H\left(T_{m+\Delta m}-t \right)H(t-y)\delta\left(t-y-T_m \right)\right) \nonumber \\
		&=\frac{1}{s+u}\mathbb{E}\left( e^{-s(\tau_0+\ldots+\tau_{m/\Delta m-1})}(1-e^{-(s+u)\tau_{m/\Delta m}})\right) \nonumber \\
		&=\hat{F}(\Delta m,s)^{m/\Delta m}\frac{1-\hat{F}(\Delta m,s+u)}{s+u} \nonumber \\
		&\sim e^{-w(s)m}\frac{w(s+u)}{s+u}\Delta m \label{eq:1_Backward_record_dm}
	\end{align}
	Similarly for the forward record time, 
	\begin{align}
		&\mathcal{L} \lbrace \mathbb{P}\left( E_t=y,M(t) \in [m,m+\Delta m] \right) \rbrace \nonumber  \\
		&=\iint {\rm d}t{\rm d}y e^{-uy-st}\mathbb{E}\left( \delta(y+t-T_{m}-\tau_{m/\Delta m})H(t-T_m) \right) \nonumber \\
		&=\mathbb{E}\left( e^{-sT_m} \frac{e^{-u\tau_{m/\Delta m}}- e^{-s\tau_{m/\Delta m}}}{s-u} \right)  \nonumber \\
		&=\hat{F}(\Delta m,s)^{m/\Delta m} \frac{\hat{F}(\Delta m,u)-\hat{F}(\Delta m,s)}{s-u} \nonumber \\
		&\sim e^{-w(s)m}\frac{w(s)-w(u)}{s-u} \Delta m \label{eq:forward_record_Laplace_dm}
	\end{align}
	This emphasizes that all formulae given in the main text are exact for the continuous processes by taking the discretization (taken to $\Delta m=1$ in the main text) infinitesimal.
	\section{Number of records, forward and backward record times}
	\label{app:MEB}
	\subsection{Single-time distribution}
	For the sake of completeness, we provide here the full derivation of joint statistics of the number of records at time $t$ as well as the backward and forward record time, $(M(t),B_t,E_t)$. To compute the distribution, we note that 
	the event $\lbrace M(t)=m,B_t=y,E_t=z \rbrace$ is exactly the event of reaching the $m^\text{th}$ record at time $t-y$ and the \blue{$(m+1)^\text{th}$} record at time $t+z$. It results in the following identity: 
	\begin{align}
		\mathbb{P}(M(t)=m,B_t=y,E_t=z)=\mathbb{E}\left( \delta\left(T_m-(t-y) \right)\delta\left(T_{m+1}-(t+z) \right)H(t-y) \right)
	\end{align}
	Once we Laplace transform this equality, we get
	\begin{align}
		&\mathcal{L}\lbrace \mathbb{P}(M(t)=m,B_t=y,E_t=z) \rbrace \nonumber \\
		&=\mathbb{E}\left( \sum_{t,y,z=0}^\infty   e^{-st-uy-vz} \delta\left(T_m-(t-y) \right)\delta\left(T_{m+1}-(t+z) \right)H(t-y) \right) \nonumber \\
		&=\mathbb{E}\left( \frac{e^{-sT_m-v\tau_m}-e^{-sT_m-(s+u)\tau_m}}{1-e^{-s-u+v}}\right) \nonumber \\
		&=\hat{F}(s)^m \frac{\hat{F}(v)-\hat{F}(u+s)}{1-e^{-s-u+v}} \nonumber \\
		&\sim e^{-w(s)m}\frac{w(s+u)-w(v)}{s+u-v}
	\end{align}
	
	\subsection{$n$-time distribution}
	We also provide the multiple-time distribution of the number of records, backward and forward record times for $m_0=0<m_1<\ldots <m_n$:
	\begin{align}
		&\mathcal{L} \lbrace \mathbb{P}\left( M(t_k)= m_k,\; B_{t_k}=y_k, \; E_{t_k}=z_k ; 1\leq k\leq n \right) \rbrace \nonumber  \\
		=&\mathbb{E}\left( \prod_{k=1}^n \iiint_0^\infty {\rm d}t_k {\rm d}y_k {\rm d}z_k e^{-v_kz_k-u_k y_k-s_kt_k} \delta\left(t_k-y_k-T_{m_k} \right) \delta\left(t_k+z_k-T_{m_k+1} \right) \right) \nonumber \\
		&=\mathbb{E}\left( \prod_{k=1}^n e^{-s_k T_{m_k}}\frac{e^{-v_k\tau_{m_k}}-e^{-(s_k+u_k)\tau_{m_k}}}{s_k+u_k-v_k}\right) \nonumber \\
		&= \prod_{k=1}^n  \Bigg[ e^{-w\left( \sum_{j=k}^n s_j \right)(m_k-m_{k-1})}\frac{w\left(u_k+\sum_{j=k}^ns_j\right) -w\left(v_k+\sum_{j=k+1}^n s_j \right) }{s_k+u_k-v_k}  \Bigg] \; .
		\label{eq:n_times_Backward}
	\end{align}
	
	\section{Records with aging}
	\label{app:aging}
	\subsection{Resetting RW}\label{app:reset}
	\subsubsection{Discrete setting}
	We define the survival probabilities from the first passage time distribution at $m$ starting at position $x$ with (noted with an $r$ index) and without resetting (see Secs.~\ref{sec:discrete_aBM} and \ref{sec:discrete_aRTP}), 
	\begin{align}
		S^r(x,t)=\sum_{t'=t}^\infty P^r(x,t'), \label{eq:Surv_reset_discrete} \\
		S(x,t)=\sum_{t'=t}^\infty P(x,t'),  \label{eq:Surv_discrete}
	\end{align}
	where $S^r(x,t)$ is the probability that the RW which resets to $0$ with probability $1-\lambda=1-e^{-r}$ at every step and starting at position $x$ (with positive speed in the case of the RTP) has not reached position $m$ by time $t$; $S(x,t)$ the probability that the RW without resetting and starting at position $x$ (with positive speed for the RTP) has not reached position $m$ by time $t$. In terms of discrete Laplace transform, Eqs.~\eqref{eq:Surv_reset_discrete} and \eqref{eq:Surv_discrete} become
	\begin{align}
		\hat{S}^r(x,s)=\frac{1-\hat{P}^r(x,s)}{1-e^{-s}}, \label{eq:Surv_reset_Lap_discrete} \\
		\hat{S}(x,s)=\frac{1-\hat{P}(x,s)}{1-e^{-s}}.  \label{eq:Surv_Lap_discrete}
	\end{align}
	Then, we use the following renewal equation, which states that the RW under resetting starting at $x$ has not reached position $m$ either without having resetted, or having resetted at a given time $t'$ and restarted at position $0$ (with positive speed in the case of the RTP).
	\begin{align}
		S^r(x,t)= \lambda^t S(x,t)+\sum_{t'=0}^{t-1}  (1-\lambda) \lambda^{t'} S(x,t')S^r(0,t-t'-1) 
	\end{align}
	Once Laplace transformed, the renewal equation becomes
	\begin{align}
		\hat{S}^r(x,s)=\hat{S}(x,s+r)+ (1-\lambda)e^{-s} \hat{S}(x,s+r)\hat{S}^r(0,s)
	\end{align}
	This results in the two equations (taking $x=0$ and $x=m-1$),
	\begin{align}
		\hat{S}^r(0,s)&=\frac{\hat{S}(0,s+r)}{1-(1-\lambda)e^{-s}\hat{S}(0,s+r)} \; ,\\
		\hat{S}^r(m-1,s)&=\hat{S}(m-1,s+r)\left(1+ (1-\lambda)e^{-s}\frac{\hat{S}(0,s+r)}{1-(1-\lambda)e^{-s}\hat{S}(0,s+r)}\right)
	\end{align}
	From Eqs.\;\eqref{eq:Surv_reset_Lap_discrete} and \eqref{eq:Surv_Lap_discrete}, we get the discrete Laplace transform of the $m^\text{th}$ record age distribution $\hat{F}^r_m(s)$, 
	\begin{align}
		1-\hat{F}^r_{m-1}(s)&=1-\hat{P}^r(m-1,s) \nonumber \\
		&=(1-e^{-s})\frac{1-\hat{P}(m-1,s+r)}{1-e^{-s}+e^{-s}(1-\lambda)\hat{P}(0,s+r)} 
	\end{align}
	Finally, using Eqs.~\eqref{eq:F_aBM_discrete} and \eqref{eq:F_aRTP_discrete}, we deduce the Laplace transform of the record age of a discrete resetting aRTP and aBM,
	\begin{align}
		1-\hat{F}_{m-1}^r(s)&= (1-e^{-s})\frac{1-\hat{F}(s+r)}{1-e^{-s}+e^{-s}(1-\lambda)\hat{F}(s+r)^m}
	\end{align}
	
	\subsubsection{Continuous setting}
	Here, we reproduce the steps described in \cite{Evans:2011,Evans:2018}. We define the survival probabilities from the first passage time distribution at $m$ starting at position $x$ with (noted with an $r$ index) and without resetting (see Secs.~\ref{sec:continuous_aBM} and \ref{sec:continuous_aRTP}),
	\begin{align}
		S^r(x,t)=\int_t^\infty F^r(x,t'){\rm d}t' \label{eq:Surv_reset} \\
		S(x,t)=\int_t^\infty F(x,t'){\rm d}t'  \label{eq:Surv}
	\end{align}
	where $S^r(x,t)$ is the probability that the RW under resetting at $0$ (with positive speed in the case of the RTP) at rate $r$ and starting at position $x$ (with positive speed in the case of the RTP) has not reached position $m$ by time $t$; $S(x,x)$ the probability that the RW without resetting and starting at position $x$ (with positive speed for the RTP) has not reached position $m$ by time $t$. In terms of Laplace transform, Eqs.\; \eqref{eq:Surv_reset} and \eqref{eq:Surv} become
	\begin{align}
		\hat{S}^r(x,s)=\frac{1-\hat{F}^r(x,s)}{s} \label{eq:Surv_reset_Lap} \\
		\hat{S}(x,s)=\frac{1-\hat{F}(x,s)}{s}  \label{eq:Surv_Lap}
	\end{align}
	Then, we use the following renewal equation, which states that the RW under resetting starting at $x$ has not reached position $m$ either because it has not reset, or it has at a given time $t'$ and restarted at position $0$ (with positive speed in the case of the aRTP).
	\begin{align}
		S^r(x,t)=e^{-rt}S(x,t)+\int_0^t r{\rm d}t' e^{-rt'}S(x,t')S^r(0,t-t') 
	\end{align}
	Once Laplace transformed, the renewal equation becomes
	\begin{align}
		\hat{S}^r(x,s)=\hat{S}(x,r+s)+ r \hat{S}(x,r+s)\hat{S}^r(0,s)
	\end{align}
	This results on the two equations (taking $x=0$ and $x=m-1$),
	\begin{align}
		\hat{S}^r(0,s)&=\frac{\hat{S}(0,s+r)}{1-r\hat{S}(0,s+r)} \; ,\\
		\hat{S}^r(m-1,s)&=\hat{S}(m-1,s+r)\left(1+r\frac{\hat{S}(0,s+r)}{1-r\hat{S}(0,s+r)}\right)
	\end{align}
	From Eqs.\;\eqref{eq:Surv_reset_Lap} and \eqref{eq:Surv_Lap}, we get
	\begin{align}
		1-\hat{F}_{m-1}^r(s)=1-\hat{F}^r(m-1,s)=s\frac{1-\hat{F}(m-1,s+r)}{s+r\hat{F}(0,s+r)}
	\end{align}
	Finally, using the expressions Eqs.~\eqref{eq:F_aBM} and \eqref{eq:F_aRTP}, we obtain Eq.~\eqref{eq:record_age_reset} of the main text, 
	\begin{align}
		1-\hat{F}_{m-1}^r(s)=s\frac{1-e^{-w(s+r)}}{s+re^{-w(s+r)m}}\sim s\frac{w(s+r)}{s+re^{-w(s+r)m}}  \equiv g(s,m)
	\end{align}
	
	\subsection{Reflecting boundary condition} \label{app:reflecting}
	We consider RWs with a reflecting boundary condition at initial position. In this case, the record age $\tau_m$ depends on $m$ as the reflecting boundary gets further and further away of the record position. For the simpler symmetric case where $\mu=0$, one can obtain the $m^\text{th}$ record age by solving Eqs.~\eqref{eq:Back_BM} and \eqref{eq:Back_RTP} on the domain $[-m,m]$ (using mirror symmetry at $x=0$) starting at position $m-1$ and absorbing boundary conditions at both ends of the interval. This results in the following expressions, 
	\begin{align}
		\hat{p}(x,s)&= \frac{\sinh\left(\sqrt{s/D} (2m-x)\right)+\sinh\left(\sqrt{s/D} x\right)}{\sinh\left(\sqrt{s/D} 2m\right)} & \text{ (BM) }\\
		\hat{p}_+(x,s)&=\frac{s T \sinh \left(\frac{x \sqrt{s \left(s+\frac{1}{T}\right)}}{v}\right)+\sqrt{s T (s T+1)} \cosh \left(\frac{x \sqrt{s \left(s+\frac{1}{T}\right)}}{v}\right)}{s T \sinh
			\left(\frac{m \sqrt{s \left(s+\frac{1}{T}\right)}}{v}\right)+\sqrt{s T (s T+1)} \cosh \left(\frac{m \sqrt{s \left(s+\frac{1}{T}\right)}}{v}\right)} & \text{ (RTP) }
	\end{align}
	which provide the following expressions of $\hat{F}_{m-1}(s)=\hat{p}(s,m-1)$ and $g(s,m)$,
	\begin{align}
		g(s,m)&=\sqrt{s/D}\tanh\left(\sqrt{\frac{s}{D}}m \right) & \text{ (BM) }\\
		g(s,m)&=\frac{\sqrt{s \left(s+\frac{1}{T}\right)} \left(\sqrt{s T (s T+1)} \sinh \left(\frac{m \sqrt{s \left(s+\frac{1}{T}\right)}}{v}\right)+s T \cosh \left(\frac{m \sqrt{s
					\left(s+\frac{1}{T}\right)}}{v}\right)\right)}{v \left(s T \sinh \left(\frac{m \sqrt{s \left(s+\frac{1}{T}\right)}}{v}\right)+\sqrt{s T (s T+1)} \cosh \left(\frac{m
				\sqrt{s \left(s+\frac{1}{T}\right)}}{v}\right)\right)} & \text{ (RTP) }
	\end{align}
	
	\twocolumngrid
	
	%\bibliography{ref}
	%\bibliographystyle{apsrev4-1}

\end{document}